\documentclass[prl,twocolumn,aps,showpacs]{revtex4}
\usepackage{amssymb}
\usepackage{graphicx}
\usepackage{graphics}
\usepackage{color}

\begin{document}

\title{Extreme events in  two dimensional disordered nonlinear lattices}
\author{A. Maluckov$^{1}$, N. Lazarides$^{2,3}$, G. P. Tsironis$^{2,3}$,
and Lj. Had\v{z}ievski$^{1}$
}
\affiliation{
$^1$Vin\v ca Institute of Nuclear Sciences, University of
    Belgrade, P. O. Box 522, 11001 Belgrade, Serbia\\
$^2$Department of Physics, University of Crete, P. O. Box 2208,
    71003 Heraklion, Greece \\
$^3$Institute of Electronic Structure and Laser,
    Foundation for Research and Technology-Hellas, P.O. Box 1527,
    71110 Heraklion, Greece
}

\begin{abstract}
Spatiotemporal complexity is induced in a two dimensional
nonlinear disordered lattice through the modulational instability
of an initially weakly perturbed excitation. In the course of
evolution we observe the formation of transient as well as
persistent localized structures, some of which have extreme
magnitude.  We analyze the statistics of occurrence of these
extreme collective events and find that the appearance of
transient extreme events is more likely in the weakly nonlinear
regime. We observe a transition in the extreme events recurrence
time probability from exponential, in the nonlinearity dominated
regime, to power law for the disordered one.
\end{abstract}

 \maketitle

Nonlinear lattices form a unique laboratory where both coherent
and incoherent processes appear simultaneously and affect their
dynamical properties \cite{Flach}. They constitute prototypical
spatiotemporally complex systems that can be studied both
theoretically and experimentally giving a wealth of information on
the physics of spatially extended complex systems. The additional
presence of quenched disorder introduces a mechanism for local
symmetry breaking that has consequences in the long-time dynamics
\cite{disorder}. One feature of selforganization of the lattices
relates to the possibility for occurrence of extreme events (EE),
i.e. the generation of transient or persistent structures with
large amplitudes that are statistically not very significant. In
the case of ocean waves, events of this type are rogue or freak
waves, i.e. waves of considerable hight that "appear from nowhere
and  disappear without a trace" \cite{Akhmediev2009a}, usually in
relatively calm seas \cite{Albeverio2006,Kharif2009Debate2010}.
Ocean rogue waves are attributed to self-focusing effects
following the development of Benjamin-Feir
instability\cite{White1998}.  Extreme waves of this type have also
been reported in nonlinear optical fibers
\cite{Solli2007,Arecchi2011}, optical cavities \cite{Montina2009},
superfluids \cite{Ganshin2009}, microwave scattering
\cite{Hohmann2010}, semiconductor lasers \cite{Bonatto2011},
femtosecond filamentation \cite{Majus2011}, soft glass photonic
crystal fiber \cite{Buccoliero2011}, etc.

In the present work we focus on two dimensional discrete nonlinear
systems with disorder that have been realized experimentally
\cite{Schwartz2007} and address the question of the conditions for
generation of very large, i.e. extreme, discrete wave events in
the process of spatiotemporal evolution.  While the production of
EEs is mediated by modulational instability (MI) the subsequent
evolution shows a complex dynamics that leads to generation as
well as destruction of waves of large magnitude. This dynamics is
directly induced by the interplay of nonlinearity and disorder; we
are seeking the regimes where EE appearance is the most frequent
as well the statistics of their appearance. In recent work we
found that integrability properties of the lattice do play a role
in the generation of EEs \cite{Maluckov2009}. The present work,
focusses on the issue of discrete lattice EEs from a broader
perspective and through a physically realizable model, viz. that
of the Anderson disordered Discrete Nonlinear Schr\"odinger (DNLS)
equation \cite{Schwartz2007,Lahini2008}.  We address two experimentally relevant questions, viz.
what is the optimal regime for EE generation and what is the
recurrence time statistics of these events once generated.  From
this analysis a clearer picture of the spatiotemporal complexity
of the system emerges.

We consider the dynamics of a strongly coupled excitation in a two
dimensional tetragonal lattice with diagonal disorder, or,
equivalently, wave propagation in a two dimensional array of
evanescently coupled optical nonlinear fibers with random index
variation, both described through the disordered DNLS equation,
viz.
\begin{widetext}
\begin{equation}
\label{1}
i\frac{d{\psi}_{n,m}}{dt}=  {\epsilon}_{n,m} {\psi}_{n,m}+J({\psi}_{n+1,m}+{\psi}_{n-1,m}+
{\psi}_{n,m+1}+{\psi}_{n,m-1})
 +\gamma \left\vert {\psi}_{n,m}\right\vert ^{2}{\psi}_{n,m}
 ,
\end{equation}
\end{widetext}
where $n,m=1,...,N$, $\psi_{n,m}$ is a probability (or wave)
amplitude at site $(n,m)$ , $J > 0$ is the inter-site coupling
constant accounting  for tunnelling between adjacent sites of the
lattice (corr. evanescent coupling), $\gamma > 0$ is the
nonlinearity parameter that stems from strong electron-phonon
coupling (corr. Kerr nonlinearity), while ${\epsilon}_{n,m}$, is
the local site energy (related to the fiber refractive index),
chosen randomly from a uniform, zero-mean distribution in the
interval $[-W/2, +W/2]$. Equation (\ref{1}) serves as a paradigmatic
model for a wide class of physical problems where both disorder
and nonlinearity are present. For $\gamma \rightarrow 0$,
Eq.(\ref{1}) reduces to the 2D Anderson model while in the absence
of disorder (${\epsilon}_{n,m}=0$), it reduces to the DNLS
equation in 2D that is generally non-integrable.  Equation
(\ref{1})  conserves the norm
   $P_N=\sum_{n=1}^N \sum_{m=1}^M |{\psi}_{n,m}|^2$,
and the Hamiltonian $\cal{H}$, corresponding to total probability
(corr. input power)  and the energy of the system, respectively.
We will consider the evolution of an initially uniform state that
is slightly modulated by periodic perturbations in order to
facilitate the development of MI \cite{Kivshar1994},
\cite{Wollert2009}. Specifically, we use the initial state
${\psi}_{n,m}=({\psi}_0 +\delta {\psi}_{n,m})\exp({-i\mu t})$,
where ${\psi}_0$ is a constant amplitude of the continuous wave
(CW) (averaged over disorder, with $\langle
\epsilon_{n,m}\rangle=0$), and $\delta {\psi}_{n,m} \equiv
a_{n,m}+ib_{n,m}$ are small complex perturbations that modulate
the constant CW solution. The MI induces nonlinearly localized
modes that are, however, modified by the presence of the quenched
disorder.  In the time scale of the numerical study, we note that
the presence  of disorder induces some form of additional energy
redistribution among the sites of nonlinear lattice.

\begin{figure}[t!]
\center
\includegraphics[width=7.5cm]{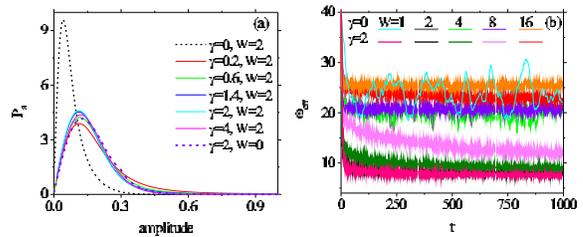}
\caption{(color online) Probability amplitude distribution $P_s$
as a function of amplitude $|\psi |$ for different nonlinearity
strengths and $W=2$ (a). The curves have a Rayleigh-like behavior.
(b) Effective average width of excitations formed in the lattice
as a function of time for different disorder levels  for $\gamma
=0,~2$. We note that for fixed nonlinearity, increase in the
amount of disorder induces delocalization.} \label{fig1}
\end{figure}

We adopt the criterion employed in oceanography and define as an
extreme wave one that has height $H_{ext}$ (here the positive
difference of minimum-to-maximum amplitude $|\psi_{n,m}|$  in
different time steps) with $H_{ext} \gtrsim 2.2 H_s$, where $H_s$,
is the significant wave height, i.e. average height of the
one-third highest waves in a given sample
\cite{Kharif2009Debate2010}. We integrate Eqs. (\ref{1}) with
periodic boundary conditions using a $6-$th order Runge-Kutta
\cite{Maluckov2009} solver for several different values of $W$ and
$\gamma$ ($J=1$) using typically a lattice with $N=41$  (we have
checked results for $N=81$ as well) and periodic boundary
conditions. During relatively long time (typically we used maximum
time of $1\times 10^3$ units or  equivalently approximately $500$
coupling lengths)
 the system selforganizes
and we observe  localized structures appearing on different sites
surrounded by irregular, low-amplitude background. Some of these
structures are in the form of breathers, either pinned or mobile,
while some others are  transient. The complete amplitude
statistics for the observed time interval is shown in Fig.
(\ref{fig1}a) for different disorder and nonlinearity parameters;
we observe a distinct, Rayleigh-like distribution that vary
depending on the parameters. Any of the states, longer lived or
transient, that appear in tail of the distribution are
 candidates for extreme events.

In order to quantify the onset of extreme events in the lattice we
use several measures, viz. the inverse participation ratio, the
probability for extreme event occurrence as well as first
appearance and recurrence EE times. The inverse participation
ratio
\begin{equation}
\label{3}
   P={P_N^2} \left\{ \sum_{n=1}^{N}\sum_{m=1}^{M}|f_{n,m}|^4 \right\}^{-1} ,
\end{equation}
determines the effective confinement of  initially extended excitations.  We define
  $\omega_{eff}=P^{1/2}$; this effective width gives the average
spatial extent of the structures generated  after the MI (Fig. (\ref{fig1}b)).

Both disorder and nonlinearity, when each acts alone, favor wave
localization in the lattice. When they are present simultaneously,
quenched disorder dominates the early dynamics  since MI develops more slowly,
at least for relatively small $\gamma$'s. In this regime, where both nonlinearity and
disorder are present, the
Anderson-like localized states decay spatially in a fashion that
still permits local energy redistribution, while at relatively longer times a more stable
state is reached.  We note from Fig. (\ref{fig1}b) that
$\omega_{eff}$ saturates to lower values for non-zero nonlinearity
compared to the  value in the corresponding linear disordered
system.  This tendency is compatible with the findings of Schwartz
et al.
 who observe experimentally that increased
self-focusing enhances localization \cite{Schwartz2007}. On the other hand, we also
see in the same figure that increasing disorder for fixed
nonlinearity results to larger $\omega_{eff}$, leading thus to
reduction of localization.  This opposite tendency of
delocalization induced by disorder in the presence of nonlinearity
can be attributed to the partial destruction of the pinned, highly
localized  nonlinear states by the presence of disorder and the
associated weak redistribution of energy.

\begin{figure}[h]
\center
\includegraphics [width=7.5cm]{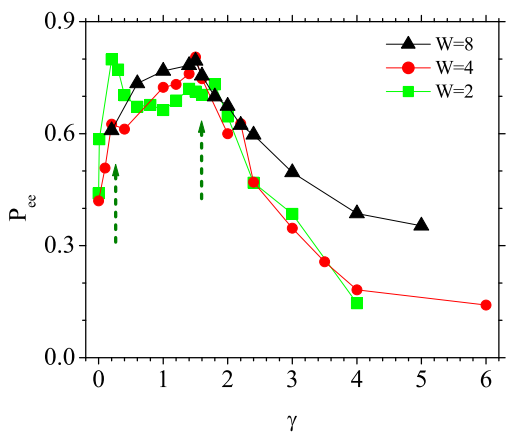}
\caption{(color online) Extreme event height probability $P_{ee}$
as a function of the nonlinearity strength $\gamma$ for several
levels of disorder. Arrows show positions of local probability
maximums.} \label{fig2}
\end{figure}

In order to find the regimes that favor EE production we calculate
numerically $P_{ee}$, the probability for EE generation with
$h>H_{ext} \equiv 2.2 H_s$; it  is shown in Fig. \ref{fig2} as a
function of $\gamma$ for three different levels of disorder. From
Fig. \ref{fig2} we observe that increasing the strength of
nonlinearity results in an increase in the EE probability $P_{ee}$,
reaching generally a maximum, while at high nonlinearity parameter
values ( $\gamma>2$ ), the EE probability reduces for all observed
disorder levels. We note that for small disorder the curve has
additionally a secondary maximum while the occurrence of the
primary maximum value shifts to large $\gamma$-values for larger
amounts of disorder $W$. Further increase of nonlinearity leads to
the decay of the extreme event generation probability, however,
with a tail that is substantially higher for the larger disorder
case. For vanishingly small nonlinearity, $P_{ee}$ is small but with
appreciable values; we find $P_{ee}=0.47,~0.44$ for disorder levels
with $W=2,~4$, respectively. These results indicate that the most
favorable conditions for the EE generation during propagation
through in the 2D lattice systems are relatively high level of
disorder compared to the linearized bandwidth and presence of
nonlinearity. The interplay of disorder and nonlinearity results
in one or more maxima in the probability of EE generation for
certain values of the nonlinearity strength. The first maximum
which is the most pronounced one for relatively small disorder and
weak nonlinearity can be correlated with high EEs probability in
the near integrable system pointed out in the 1D context
\cite{Maluckov2009}. On the other hand, for arbitrary disorder
the localizing effect of
nonlinearity  seems to be responsible for the appearance of the second
maximum at $\gamma \approx 2$ with a value  $P_{ee}\approx 0.7\% $.

\begin{figure}[h]
\center
\includegraphics[width=7.5cm]{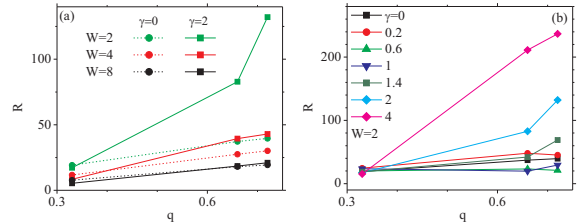}
\caption{(color online) The average return time $R$ of extreme
events as a function of  the threshold amplitude $q$ for (a)
various disorder levels with $\gamma =0$ (circles) and $\gamma =2
$, (b) fixed level of disorder ($W=2$)  and several nonlinearity
strengths (including zero nonlinearity). } \label{fig3}
\end{figure}

In order to probe deeper into the genesis and appearance
statistics of EEs in discrete, disordered nonlinear lattices we
focus on the probability distribution $P_r$ of the recurrence time
$r$ of the extreme events \cite{Altmann2005}, \cite{Santhanam2008}.
We study two quantities, the mean recurrence time $R$ of an EE as
well as the mean time of first appearance (TFA).  As these times
depend on the specific value of the parameter $q$, the wave height
threshold, we first investigate  their dependence on $q$.
Numerics show that the slope of the curves  of the recurrence time
$R$ as a function of the threshold $q$  is smaller when only
disorder is present compared to the case of simultaneous  presence
of disorder and nonlinearity for all levels of disorder (Fig.
\ref{fig3}a). The mean recurrence time $R$ increases with $q$ for
all values of $W$ and $\gamma$ (Fig. \ref{fig3}). The increase is
faster for lower disorder level (Fig. \ref{fig3}a) and stronger
nonlinearity (Fig. \ref{fig3}b). This dependence in the strong
nonlinearity-weak disorder regime is related to the favorable
conditions  for the creation of highly pinned, immobile localized
structures, a fact that increases dramatically the mean recurrence
time.
Qualitatively similar behavior we found also for the TFA-$q$
dependence.  These studies show the complexity of the recurrence
as well as first appearance phenomena and their direct
manifestation in time average quantities.

\begin{figure}[h]
\center
\includegraphics[width=7.5cm]{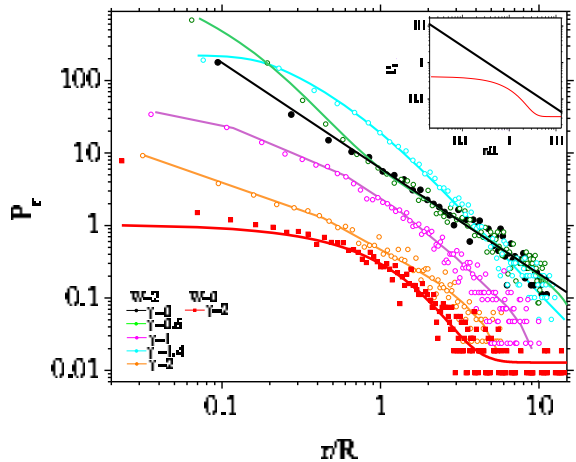}
\caption{(color online) Return time probability $P_r$ (not
normalized) as a function of $r/R$, the return time scaled by the
corresponding average return time, for the purely disordered case
(power law fit - black curve), purely nonlinear case (exponential
fit - red line) and intermediate cases (double exponential fits).
We use $q=0.66$. In box the lines which designate the first two
cases are shown separately.} \label{fig4}
\end{figure}

We calculate $P_r$, the return time probability for different
disorder levels and nonlinearity values in the following way.  For
a given threshold value $q$ ($q>H_{ext}$) we scan the lattice and
find an event at a given location with amplitude larger than $q$;
we register as recurrence time  the time between this event and a
subsequent above threshold  one that appears in the same lattice
location .  We follow this procedure repeatedly up to maximum time
and construct histograms for different parameter values, all
scaled by the average return time $R$ in each parameter regime;
the outcomes for the various non normalized return probabilities
are shown in  Fig. \ref{fig4} for several values of disorder level
and nonlinearity for a given $q$-value. In the purely linear but
disordered regime we find that  $P_r$ is a  power-law function,
viz. $P_r=[a+b(r/R)]^{-\beta}$, with $\beta \approx 1.34$
$q=0.66$;  while the parameters depend on the threshold value $q$,
the functional form is stable. The addition of nonlinearity to a
disordered lattice reduces somehow the   possibility of occurrence
of  EEs. Remarkably, the $P_r$ vs. $r/R$ dependence of the curves
changes gradually from power-law, in the linear, disordered
regime, to a double  exponential at intermediate nonlinearity
values to single exponential for a pure nonlinear lattice with no
or very little disorder. Furthermore, the  decay of the curves is
more rapid for higher nonlinearity strength.  The observed transition
from a power law, in the disordered dominated regime, to exponential,
in the nonlinearity dominated one, is linked to the behavior of the
tail of the amplitude probability distributions.

\begin{figure}[h]
\center
\includegraphics[width=7.5cm]{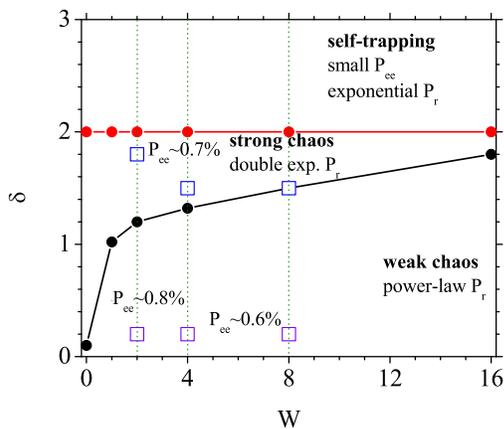}
\caption{(color online) Different regimes of the spreading of
wave-packets in the effective parameter space $(W,\delta)$
~\cite{Flach1d,Flach2d}. Lines represent regime boundaries
$\delta\approx d$ and $\delta =2$. Empty squares denote parameters
for which the local maximums of $P_{ee}$ are found (\ref{fig2}). }
\label{fig5}
\end{figure}

We relate our relatively short-time results to expected long-time
wave-packet spreading regimes summarized in Ref. \cite{Flach1d}
and \cite{Flach2d} (one-dimensional DNLS with disorder and its
generalization to the higher dimensional problem, respectively).
Different regimes are obtained for (i) $\delta > 2$, onset of
selftrapping, (ii) $d<\delta <2$, strong chaos and (iii) $\delta <
d$ for the effective quantities $d$ and $\delta$. The latter are
the average frequency spacing of the nonlinear modes within a
localization volume given by  $d\approx
\Delta/V=(8J+W)/\omega_{eff}$ while  the nonlinear frequency shift
$\delta$ is taken to be $\delta\approx \gamma$. The selection of
regimes in \cite{Flach1d, Flach2d} was done taking into account
the intensity of interaction among the nonlinear modes; the latter
increases with the nonlinearity strength up to the high
nonlinearity (here $\delta\approx 2$) when the strong
self-trapping results in a creation of isolated strongly pinned
high amplitude breathers. The comparison is  approximate but leads
to
 interesting observations (see Fig. \ref{fig5}). The
first local maximum in the $P_{ee}$ found for small nonlinearity
strength is located in the weak chaos regime. Its maximal value
$P_{ee}\approx 0.8\% $ is observed for small $W$. The second, broader
maximum of $P_{ee}$ for all disorder cases is located  in the strong chaos
regime relatively close to the border lines with neighboring
regimes (Fig. \ref{fig5}). On the other hand,  we may associate
the power-law decay of the $P_r$ to the weak chaos regime, and the
exponentially like one with the self-trapping regime. Therefore,
transient EEs are more probable in the regime of weak chaos, while
the long-lived EEs (high amplitude strongly pinned breather
structures) are dominant structures in the regimes of strong chaos
and self-trapping. This association enables us to relate the
first local maximum in $P_{ee}$ to the weak interaction of
transient EEs induced by disorder while the second, broader maximum, to the
appearance of longer-lived breathers resulting from  the energy
redistribution through the strong interaction between nonlinear modes.

We have calculated several statistical measures for wave propagation in
disordered and nonlinear lattices described by the DNLS in two-dimensions
with respect to EEs generation. The interplay of two distinct mechanisms
of energy localization leads to several unexpected results.
The calculations indicate that EEs may be generated with appreciable
probability also in linear disordered lattices (but not in linear non-disordered lattices)
for the chosen type of initial excitations.

When the relative strengths of nonlinearity and disorder vary, the
average return probability $P_r$ of EEs as a function of $r/R$
changes between the two limiting regimes, i.e., disorder without
nonlinearity and nonlinearity without disorder corresponding
respectively to a transition  from power-law to exponential .
These  findings are illustrated in Fig. \ref{fig4} for certain
parameter values.  The change occurs gradually between the two
regimes passing through a sequence of  states that have more
complex return time behavior. Moreover, the calculated probability
for EE generation indicates that there are two parameter regimes
of the disorder level and nonlinearity strength which favor EE
generation, as shown in Fig. (\ref{fig2}). We found that a
relatively low level of disorder (compared to the bandwidth) leads
to maxima in the probability for EE production  in the
 relatively weak nonlinearity regime. The second, broader, maximum found for higher nonlinearity
 does not depend significantly on the disorder level. For very high
$\gamma$'s the probability of EE generation, although small,
depends strongly on the level of the disorder. The fact that EE
generation is favorable in the weakly nonlinear regime appears to
be compatible
 with previous findings in discrete lattices, where
the highest probability for EE generation is observed in the
nearly integrable limit \cite{Maluckov2009}. On the other hand,
the appearance of the second parameter regime which favors the EE
generation is correlated with the localization mechanism governed
by nonlinearity \cite{Molina1993}. We also note that the mechanism responsible for
the  probability for EE generation depends additionally on
the specific initial conditions used, as seen also in other
approaches \cite{Bludov2009}.

\acknowledgments A.M. and Lj.H. acknowledge support from the
Ministry of Education and Science, Serbia (Project III45010).
G.P.T and N.L. acknowledge partial support of the Thalis project
MACOMSYS of the Greek Ministry of Education.

\end{document}